\documentclass{cernrep}
\begin{document}
\title{ Anomalous gauge interactions in photon collisions at the LHC and the FCC 
 }
\author{S. Fichet$^a$\footnote{Speaker}, C. Baldenegro$^b$     }
\institute{  $^a$ ICTP-SAIFR \& IFT-UNESP, R. Dr. Bento Teobaldo Ferraz 271, S\~ao Paulo, Brazil \\
$^b$ University  of Kansas, Lawrence, Kansas, U.S. }

\begin{abstract}

The  forward proton detectors recently installed and operating at the LHC open the possibility to observe photon collisions with high precision, providing a novel window on physics beyond the Standard Model. We review recent simulations and theoretical developments about the measurement of anomalous $\gamma\gamma\gamma\gamma$ and $Z\gamma\gamma\gamma$ interactions. 
The searches for these anomalous gauge interactions are expected to set bounds on a wide range of  particles including  generic electroweak  particles, neutral particles with dimension-5 coupling to gauge bosons, polarizable dark particles, and are typically complementary from new physics searches in other channels.

\end{abstract}


\maketitle

\newcommand{\be}{\begin{equation}}  
\newcommand{\ee}{\end{equation}}  
\newcommand{\bea}{\begin{eqnarray}}  
\newcommand{\eea}{\end{eqnarray}}

\section{ Forward proton detectors and photon collisions}

\begin{figure}
\begin{center}
\includegraphics[trim=0cm 0cm 0cm 0cm, clip=true,width=12cm]{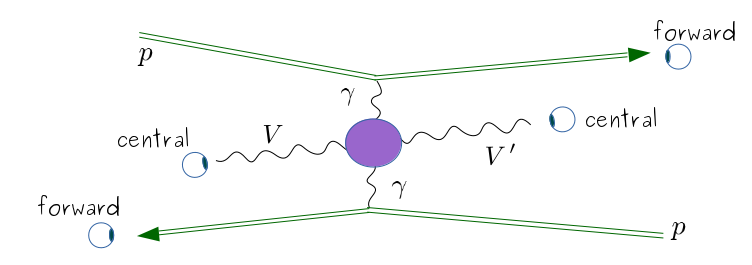}
\end{center}
\caption{Central exclusive photon collision with gauge boson final states. The protons remain intact and are seen by the forward detectors. The outgoing gauge bosons have high $p_T$ and are seen in the central detectors. }
\label{fig:collision}
\end{figure}

The new forward detectors have been installed at both ATLAS (ATLAS Forward Proton detector \cite{atlas}) and CMS (CT-PPS detector \cite{cms}) and have started to take data since the 2017 run.
The purpose of these detectors is to measure intact protons arising from diffractive processes at small angle, giving access to the so-called central exclusive processes  
\be
pp\rightarrow p \,\oplus X  \,\oplus p\,,
\ee
where the $\oplus$ denote gaps with no hadronic or electromagnetic activity between the central system $X$ and the outgoing protons, see Fig.~\ref{fig:collision}. 
These central exclusive processes can potentially provide a new window on physics beyond the Standard Model (SM) at the LHC.  The special role of the forward detectors is to characterize the outgoing intact protons, hereby giving access to  the complete kinematics of the event. Such information can then be used to drastically reduce the backgrounds.

 The forward detectors are installed at $\sim200$ m on both sides of CMS and ATLAS and host  tracking stations. 
 Their acceptance in the fractional momentum loss of the intact protons  $\xi$ is approximately $0.015 < \xi < 0.15$ at the nominal accelerator magnetic lattice and beam conditions, which corresponds to an acceptance of  $300$ to $1900$ GeV in the invariant mass of the central system for the nominal LHC beam optics when both protons stay intact.  
 The CMS and TOTEM collaborations presented the first results of said system at the LHC by measuring the central semi-exclusive production of high-mass muon pairs at 13 TeV with an integrated luminosity of 10 fb$^-1$ collected in high-luminosity fills \cite{CMS:2017uey}, which proves the feasibility of the search for New Physics in the exclusive channel.   
Timing detectors are scheduled to be installed  in both  CT-PPS and ATLAS to measure the protons time-of-flight of protons with an expected precision of $\sim$ 15 ps, which would allow to determine the primary vertex with a $\sim1$ mm precision\,\cite{Timing}. Timing detectors will not be used in the estimations presented today, however they are absolutely necessary for final states with missing transverse energy (for example, $W^+W^-$ in the leptonic channel).


 Central exclusive processes with intermediate photons -- \textit{i.e.} photon collisions -- are especially interesting as the equivalent photon flux from the protons is large and well understood.  
In terms of an effective theory description of the new physics effects, one can for instance test operators like $|H|^2 F_{\mu\nu}F^{\mu\nu}/\Lambda^2$ which induce anomalous  single or double Higgs production (for the MSSM case, see \cite{Heinemeyer:2007tu, Tasevsky:2014cpa}). Our focus is on quartic gauge couplings, which leads to diboson final states as shown in  Fig.~\ref{fig:collision}.  Such self-interactions of neutral gauge bosons are particularly appealing to search for new physics, because the SM irreducible background is small. The neutral quartic gauge interactions constitute smoking gun observables for new physics.
Studies using proton-tagging at the LHC 
for new physics searches can be found in \cite{usww, usw,Sahin:2009gq,Atag:2010bh, Gupta:2011be, Epele:2012jn, Lebiedowicz:2013fta, Fichet:2013ola,Fichet:2013gsa,Sun:2014qoa,
Sun:2014qba,Sun:2014ppa,Sahin:2014dua,Inan:2014mua,Fichet:2014uka,Fichet:2015nia,Cho:2015dha,Fichet:2016clq}.

\label{se:fwd}

\section{ Searching for anomalous $\gamma\gamma\gamma\gamma$ and $Z\gamma\gamma\gamma$ interactions }

Given the promising possibilities of forward detectors,  realistic simulations of the search for $\gamma\gamma \rightarrow  \gamma\gamma$, $\gamma\gamma \rightarrow Z  \gamma $  have been carried out in \cite{Fichet:2013gsa, Fichet:2014uka, Baldenegro:2017aen}. 
The search for light-by-light scattering at the LHC without proton tagging  has been first tho\-roughly analyzed in \cite{dde:2013yra}. Let us review the setup, the backgrounds and the event selection for $\gamma\gamma$ and $Z\gamma$ final states.


The Forward Physics Monte Carlo  generator (FPMC, \cite{FPMC})  is designed to produce within a same framework the double pomeron exchange (DPE), single diffractive,
exclusive diffractive and photon-induced processes. 
The emission of photons by protons is correctly described by 
the  Budnev flux \cite{Chen:1973mv, Budnev:1974de}, which takes into account the proton electromagnetic structure.
The SM $\gamma\gamma \rightarrow  \gamma\gamma$  process induced by loops of SM fermions and $W$ boson, the exact contributions from new particles with  arbitrary charge and mass, and 
 the anomalous vertices described  by the effective operators of Eq.~\eqref{eq:zetas} have been implemented into FPMC.

The main background for the signals $\gamma\gamma \rightarrow  \gamma\gamma, Z(ll,jj) \gamma$ is the detection of central final states  (including misidentified jet  or electrons) originating from a non-exclusive process, occurring \textit{simultaneously} with the forward detection of two protons  from pile-up. The probability to detect at least one proton in each of the forward detectors is estimated to be 32\%, 66\% and 93\% for 50, 100 and 200 additional interactions respectively. 
Other backgrounds include double Pomeron exchange and central exclusive QCD production. 

The knowledge of the full event kinematics is a powerful constraint
to reject the huge background from pile-up. The key requirements consist in matching the missing momentum (rapidity difference) of the di-proton system  with the invariant mass (rapidity difference) of the central system ($XY\equiv\gamma\gamma, (jj)\gamma$, $(\ell\bar{\ell})\gamma$), with $m_{XY}/m_{pp}< 5-10\%$ and $|y_{XY}-y_{pp}|<3-10\%$ depending on the channel.
 Extra cuts rely on the event topology, using the fact that the $XY$ states are typically back-to-back with similar $p_T$.  This translates to cuts of the form  $|\Delta \phi_{X,Y} - \pi|<0.02$, $p_{T X}/p_{T  Y}>0.90-0.95$. The signal is harder than the background hence one cuts on the invariant mass typically as  $m_{XY}>600-700$ GeV.  
  Further background reduction could  be possible by  measuring the protons
time-of-flight, which would allow to constrain the event vertex.

%
%
%

\section{ Sensitivity to heavy new particles at LHC and FCC}

In a scenario where new particles are too heavy to be produced on-shell at the collider, the presence of these new  states is best studied using effective field theory methods. The low-energy effects of the new particles are parametrized by higher dimensional operators made of SM fields. For the quartic gauge interactions of our interest we use the basis of operators 
\bea
\mathcal{L}_{4\gamma}&=&\zeta^\gamma F_{\mu\nu}F^{\mu\nu}F_{\rho\sigma}F^{\rho\sigma}+\,\tilde\zeta^\gamma F_{\mu\nu}\tilde F^{\mu\nu}F_{\rho\sigma} \tilde F^{\rho\sigma} \\ \nonumber
\mathcal{L}_{3\gamma Z}&=&\zeta^{\gamma Z} F_{\mu\nu}F^{\mu\nu}F_{\rho\sigma}Z^{\rho\sigma}+\,\tilde\zeta^{\gamma Z} F_{\mu\nu}\tilde F^{\mu\nu}F_{\rho\sigma} \tilde Z^{\rho\sigma}\,.\label{eq:zetas}
\eea
Other operators like ${\cal O}_2^\gamma = F_{\mu\nu}F^{\nu\rho}F_{\rho\sigma}F^{\sigma\mu}$ are sometimes used. They are linearly dependent of the ones above, with  $4{\cal O}_2 = 2{\cal O}+ \tilde{\cal O}$.

These expected sensitivities  are given in Table~\ref{tab:sensitivities}
for different scenarios corresponding to medium luminosity 
(300~fb$^{-1}$) and to high luminosity (3000~fb$^{-1}$) at the LHC. The background being small, statistical significance grows quickly with the event number. Typically $O(10)$ events are enough to reach $5\sigma$ significance. 
 We also provide an estimation for FCC with proton collisions, assuming 3000~fb$^{-1}$ and average pile-up of $\mu=200-1000$. We assume the forward proton detectors properties at FCC to be similar to those at the LHC, including the acceptance range.

\begin{table}[t!]
\caption{5\,$\sigma$ sensitivity to the effective quartic gauge couplings of Eq.~\eqref{eq:zetas} in GeV$^{-4}$. 
}

\begin{center}

\resizebox{\columnwidth}{!}{
\begin{tabular}{|c|c|c|}
\hline
 Assumptions & $\gamma\gamma$ final state  & $\gamma jj$, $\gamma ll$ final state \\
\hline 
 & & $\zeta^{\gamma Z}(\tilde\zeta^{\gamma Z})<1.9\cdot 10^{-13}$ $[\gamma jj+\gamma ll]$
\\
13 TeV, 300~fb$^{-1}$, $\mu=50$ & $\zeta^\gamma(\tilde\zeta^\gamma)<9\cdot 10^{-15}$  & $\zeta^{\gamma Z}(\tilde\zeta^{\gamma Z})<2.8\cdot 10^{-13}$ $[\gamma ll]$\\
 &   & $\zeta^{\gamma Z}(\tilde\zeta^{\gamma Z})<2.3\cdot 10^{-13}$ $[\gamma jj]$
\\
 \hline
13 TeV, 3000~fb$^{-1}$, $\mu=200$ &  $\zeta^\gamma(\tilde\zeta^\gamma)<1\cdot 10^{-14}$  &
$\zeta^{\gamma Z}(\tilde\zeta^{\gamma Z})<1.8\cdot 10^{-13}$ $[\gamma ll]$
 \\ \hline
100 TeV, 3000~fb$^{-1}$, $\mu=200$ &  $\zeta^\gamma(\tilde\zeta^\gamma)<1.1\cdot 10^{-16}$  & \\ \hline
100 TeV, 3000~fb$^{-1}$, $\mu=1000$ &  $\zeta^\gamma(\tilde\zeta^\gamma)<2\cdot 10^{-16}$  & \\ \hline
\end{tabular}
}
\end{center}

\label{tab:sensitivities}

\end{table}

The ATLAS Collaboration has set a bound on the $Z\rightarrow \gamma\gamma\gamma $ decay of 
${\cal B}(Z\rightarrow \gamma\gamma\gamma)<2.2 \cdot 10^{-6} $ \cite{Aad:2015bua},  beating the ones from LEP.  This bound translates as a limit \cite{Baldenegro:2017aen}
\be
\sqrt{\zeta^2+\tilde\zeta^2 -\frac{\zeta \tilde \zeta}{2}  } < 1.3 \cdot 10^{-9} ~\textrm{GeV}^{-4}~~ (95\%{\rm CL}) \,.
 \label{eq:ATLAS_bound} 
\ee 
Imagining the same search is done at $13$ TeV data with $300$~fb$^{-1}$ in the same conditions, we expect very roughly an improvement by an order of magnitude of the bound of  Eq.~\ref{eq:ATLAS_bound}.  In addition, the current number of pile-up interactions at $13$~TeV sets a challenge to the measurement of $3\gamma$ final states. This remains far away from the expected sensitivities obtained from photon collisions at the same luminosity by roughly three orders of magnitudes.



\vspace{4 cm}

\section{ Sensitivity to new electroweak particles}

\begin{figure}[h]
\begin{center}
\includegraphics[trim=0cm 0cm 0cm 0cm, clip=true,width=5cm]{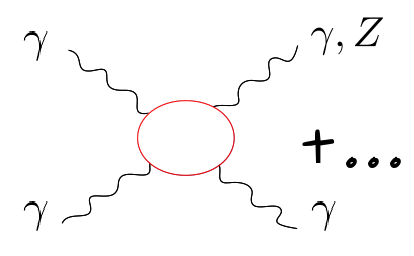}
\end{center}
\caption{ A loop of  electroweak particles inducing a quartic gauge interaction. }
\label{fig:LoopEW}
\end{figure}

\begin{figure}[h]
\begin{center}
\includegraphics[trim=0cm 0cm 0cm 0cm, clip=true,width=10cm]{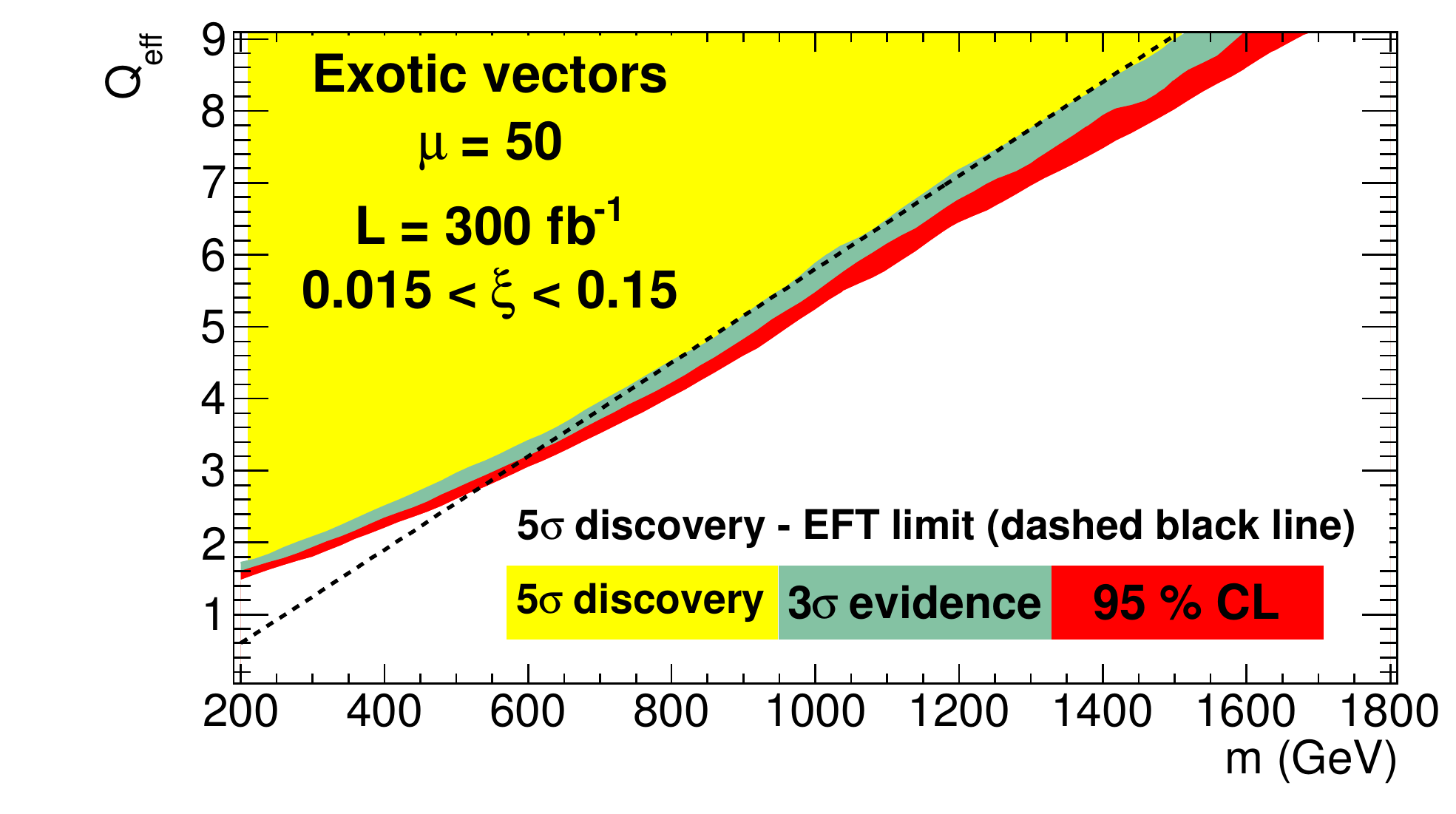}
\end{center}
\caption{ Expected sensitivity on a spin-1 particle as a function of its mass  and of its effective charge $Q_{\rm eff}$.  }
\label{fig:vector}
\end{figure}

New EW charged particles contribute to anomalous gauge couplings  at one-loop (Fig.~\ref{fig:LoopEW}). 
Because of gauge invariance, these contributions can be parametrized in terms of the mass and quantum numbers of the new particle \cite{Fichet:2013ola}.  Using the heat kernel calculation of \cite{Fichet:2013ola}, we obtain the effective couplings
\bea \label{eq:loop}
\Big(\zeta^{\gamma Z}, \tilde \zeta^{\gamma Z}\Big) &=& \Big(c_{s}, \tilde c_{s}\Big)\, \frac{ \alpha^2_{\rm em} }{ s_w c_w \, m^4 }\,N \, d \left( c_w^2 \frac{3d^4-10d^2+7}{240} +  (c_w^2-s_w^2)\frac{(d^2-1)Y^2}{4} - s^2_w Y^4 \nonumber
  \right) \\ &\equiv & \Big(c_{s}, \tilde c_{s}\Big)\, \frac{ \alpha^2_{\rm em} }{ s_w c_w \, m^4 } \frac{Q^4_{\rm eff}}{4} \,, 
\eea
\be \label{eq:loop}
\Big(\zeta^\gamma, \tilde \zeta^\gamma\Big) = \Big(c_{s}, \tilde c_{s}\Big)\, \frac{ \alpha^2_{\rm em} }{  m^4 } \,N \, d \left( \frac{3d^4-10d^2+7}{960} +  \frac{(d^2-1)Y^2}{8} + \frac{Y^4}{4}
  \right) \,. 
\ee
We have evaluated the sensitivities using the exact amplitudes. For example, the sensitivity for the case of a charged vector is shown in Fig.~\ref{fig:vector} for a luminosity of 300 fb$^{-1}$ and $\mu=50$. 


Certain candidates like the typical vector-like quarks arising in Composite Higgs models are already excluded by stronger bounds. On the other hand, our expected bounds from photon collisions are completely model-independent. They apply 
to any quantum number, are independent of the amount 
of mixing with the SM quarks, and even apply to vector-like leptons.

%


\vspace{3cm}

\section{ Sensitivity to new neutral particles}

\begin{figure}[h]
\begin{center}
\includegraphics[trim=0cm 0cm 0cm 0cm, clip=true,width=5cm]{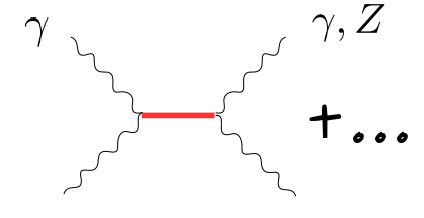}
\end{center}
\caption{ The exchange of a neutral particle inducing a quartic gauge interaction. }
\label{fig:neutral}
\end{figure}

Non-renormalizable interactions of neutral particles are 
also present in common extensions of the SM. Such theories can contain 
scalar, pseudo-scalar and spin-2 resonances, respectively denoted by $\varphi$, 
$\tilde \varphi$ and $h^{\mu\nu}$~\cite{Fichet:2013gsa}, that can  potentially be strongly-coupled to the SM. The full effective theory for such neutral resonances is  given in \cite{Fichet:2015yia}. 
\begin{equation} \begin{split} 
\mathcal L_{\gamma\gamma}= &\varphi\, 
\left[\frac{1}{f_{0^+}^{\gamma\gamma}}\, (F_{\mu\nu})^2+\frac{1}{f_{0^+}^{\gamma Z}}\, F_{\mu\nu} Z_{\mu\nu}\right]
+ \tilde\varphi  \, \left[ \frac{1}{f_{0^-}^{\gamma\gamma}}\, F_{\mu\nu} \tilde F_{\mu\nu}\,
+\frac{1}{f_{0^-}^{\gamma Z}}\, F_{\mu\nu} \tilde Z_{\mu\nu}\,
 \right] 
\\&  +  h^{\mu\nu}\, \left[ \frac{1}{f_{2}^{\gamma\gamma}} (-F_{\mu\rho} 
F_{\nu}^{\,\,\rho}+\eta_{\mu\nu} (F_{\rho\lambda})^2/4) +
\frac{1}{f_{2}^{\gamma Z}} (-F_{\mu\rho} 
Z_{\nu}^{\,\,\rho}+\eta_{\mu\nu} F_{\rho\lambda}Z_{\rho\lambda}/4)\right]\,,
\end{split}
\end{equation}
where the $f_S$ have mass dimension one.
 Such neutral particle  induces  quartic gauge couplings (Fig.\ref{fig:neutral}) given by
\be (\zeta^\gamma, \tilde \zeta^\gamma)= \frac{1}{(f^{\gamma\gamma}_{s})^2\, m^2} (d_{ s}, \tilde d_{ s}) 
\,,\quad 
 (\zeta^{\gamma Z}, \tilde \zeta^{\gamma Z})= \frac{1}{f^{\gamma\gamma}_{s}f^{\gamma Z}_{s}\, m^2} (d_{ s}, \tilde d_{ s})  
 \,,\quad
(d_{s},\tilde d_{s})=
\begin{cases}
1,0 & s=0^+ \\
 0,1 & s=0^- \\
\frac{1}{4},\frac{1}{4} & s=2 \,.\\
\end{cases}
 \ee
For example, at 13 TeV, 300 fb$^{-1}$, $\mu=50$, the expected $5 \sigma$ sensitivities on the scalar are
\be
m<4.5\, {\rm TeV}\cdot \left(\frac{1\,\rm TeV}{f^{\gamma\gamma}}\right) \,, \quad m<2.3\, {\rm TeV}\cdot \bigg(\frac{1\,\rm TeV}{\sqrt{f^{\gamma\gamma}f^{\gamma Z}}}\bigg) \,.
\ee
Following \cite{Fichet:2013gsa,Fichet:2014uka}, the bounds on a dilaton mass can reach $4260$ GeV ($5\sigma$) and those on a KK graviton with IR gauge fields can reach $5670$ GeV ($5\sigma$). 
Interestingly, the $Z\gamma $ coupling vanishes( $f_{Z\gamma}\rightarrow \infty$) when the particle couples universally to the $SU(2)\times U(1)_Y$ field strengths $(B^{\mu\nu})^2$, $(W^{I,\mu\nu})^2$, hence the $\gamma Z$ channel provides a powerful piece of information.

\section{ Sensitivity to a polarizable dark sector}

\begin{figure}[h!]
\begin{center}
\includegraphics[trim=0cm 0cm 0cm 0cm, clip=true,width=5cm]{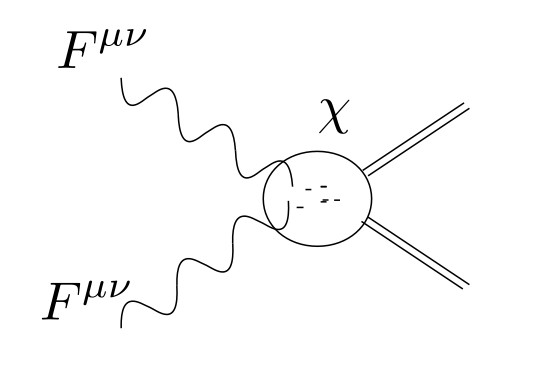}
\hspace{2.2cm}
\includegraphics[trim=0cm 0cm 0cm 0cm, clip=true,width=7cm]{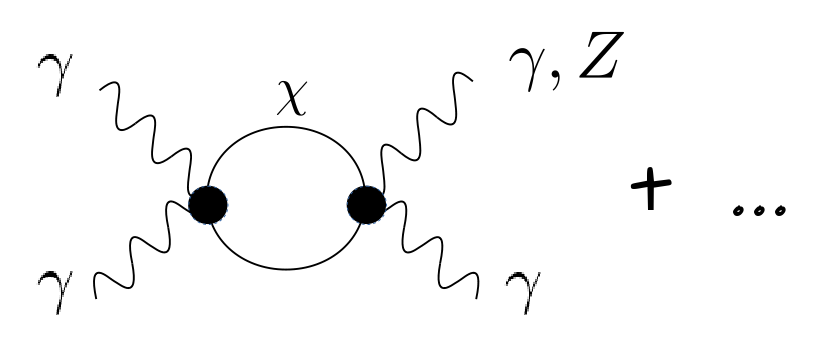}
\end{center}
\caption{\textit{Left:} Sketch of a composite polarizable dark particle. \textit{Right:} Sketch of the quartic gauge interaction induced by the polarizable dark particle.  }
\label{fig:DMpol}
\end{figure}

It is very plausible that a new particle  be \textit{almost dark} in the sense that it interacts with light only via higher-dimensional operators. Let us focus on self-conjugate particles, which have no dipole -- our  approach applies very similarly if the dark particle is polarized. For concreteness we focus on a real scalar.  Its interactions with light have the form $\frac{1}{\Lambda^2}\phi^2 (F^{\mu\nu})^2, \, \frac{1}{\Lambda^4}\partial_\mu \phi \partial_\nu \phi F^{\mu\rho}F^{\,\nu}_\rho,\, \ldots
$.  The property of polarizability can be either induced by mediators, or arise from the inner structure of the particle (intrinsic polarizability) \cite{Fichet:2016clq}. The latter happens in particular if the dark particle is a composite made of electrically charged constituents. As a matter of fact, many models of dark sectors can feature this kind of composite dark particles (see for instance vectorlike confinement, stealth DM \cite{Appelquist:2015zfa}). 

The polarizable dark particle can induce quartic gauge interactions at one-loop as shown in Fig.~\ref{fig:DMpol}.  This contribution is potentially dominant in the scenario of intrinsic polarizability. 
The photon collision then quite literally sheds light on the dark particle. In a simulation at 13 TeV, 300 fb$^{-1}$ $\mu=50$, the sensitivities in mass and in $\Lambda$ go typically up to the TeV scale -- details 
depend on which polarizability operator is considered. 

Searching for dark sectors via their virtual effects is a recent trend \cite{Fichet:2017bng,Voigt:2017vfz}, and many interesting developments are yet to be done. An advantage of such searches is they do not rely on the hypothesis of stability of the dark particle. 
When the dark particle is stable it is a dark matter candidate,  in which case the search in photon collisions turns out to be complementary to monojet+ missing energy searches. The  search in photon collisions gets favored by larger multiplicities in the loop ($\propto N^2$ versus $ \propto N$) and for large dark particle mass as it is not constrained by the kinematic threshold of DM pair production.

\section*{Acknowledgements}

I would like to thank the organizers of the Photon 2017 conference for the kind invitation and  the S\~ao Paulo Research Foundation (FAPESP) for the support under grants \#2011/11973 and \#2014/21477-2.

\end{document}